\documentstyle[12pt,fleqn]{article}
\begin{document}
\baselineskip= 22 truept
\def\be{\begin{equation}}
\def\ee{\end{equation}}
\def\bea{\begin{eqnarray}}
\def\eea{\end{eqnarray}}
\begin{titlepage}
\title{
\begin{flushright}
{\sc ip/bbsr/94-54} \\
 hep-th/9410068 \\
\end{flushright}
\vspace{4cm}
\bf Thermodynamics of Two Dimensional Black Holes}
\author{\bf Alok Kumar and Koushik Ray \\ \\
Institute of Physics, \\Bhubaneswar 751 005, INDIA \\
email: kumar, koushik@iopb.ernet.in}
\date{\today}
\maketitle
\thispagestyle{empty}
\vskip .6in
\begin{abstract}
Thermodynamic relations for a class of
2D black holes are obtained corresponding to observations
made from finite spatial distances. We also study the
thermodynamics of the charged version of the Jackiw-Teitelboim black
holes found recently by Lowe and Strominger. Our results corroborate,
in appropriate limits, to  those obtained
previously by other methods. We also analyze the
stability of these black holes thermodynamically.
\end{abstract}
\vfil
\end{titlepage}
\eject

It is customary in the study of black holes to write the laws of
thermodynamics in the asymptotic
space-time\cite{GibHaw2}. For a
static uncharged black hole in four dimensions,
the Hawking temperature is $ T_c =
\frac{1}{8 \pi M} $, where $ M $ is the mass of the black hole.
This temperature is measured at spatial infinity. Then Hawking's
interpretation of the gravitational action as the entropy leads
to the following asymptotically valid equation for this black hole:
\be \label{first}
S T_c = \frac{M}{2}.
\ee
Similar laws also exist for other black holes. However, if this
equation is true for all $ M $, then the specific heat of
the black hole is negative; thus rendering the space-time
unstable to radiation. It will lose mass with increasing
temperature and is destined to meet with a catastrophic end.
It has recently been pointed out {\cite{BY}}, that the
negativity of the specific heat of this black hole could be
tracked down to ignoring  the ${\cal O}(\frac{1}{r})$ term in the
action. Retention of this term is tantamount to
measuring the thermodynamic quantities at a finite $ r $.
Therefore one needs a finite-space formulation of black hole
thermodynamics.

Such a formulation in the context of an
effective 2D string theory has been
proposed in {\cite{GP}}. One of the major
motivations for analysing these 2D black hole solutions
in string theory is to gain some insight
into the nature of problems in {\em thoroughly un-understood}
4D theory of gravity with quantum effects.
The 2D solutions can be obtained from higher dimensional ones,
via some compactification scheme, say, and are simpler to deal with.

In string theory, as is well-known, a scalar field, namely
dilaton, plays an important role in obtaining the black hole
solutions and their physics. For example, in 2D,  pure Einstein
gravity does not have any nontrivial solutions. But the
introduction of dilaton gives rise to many interesting solutions,
such as black holes\cite{witten}, cosmological
universes\cite{venezia}  etc.
Dilaton field also plays an important role in
defining the thermodynamics of the black holes, as noted
earlier\cite{GP}.  The root of many of these new features is
the non-minimality of the
coupling of gravity and other fields to the dilaton.
Therefore, the study of other non-minimally
coupled scalars showing
up in the low energy effective action of string theory is also
of interest. Example of one such  class of 2D black holes have
been found in\cite{CM}(see also \cite{kiem}). These
come from the compactification of the 4D black holes in presence
of non-minimally coupled moduli fields in the
extremal limit. These 2D black holes asymptote a space-time of
constant negative curvature. This led to a vanishing
temperature for the corresponding
4D theory, except for the special case when the 2D black hole
corresponds to the one in string theory.
However we shall show that an analogue of equation(1) is still
valid in the 2D theory.

Introduction of an electric field in the low energy effective
field theory ensuing from sting theory has more exotic
effects.
In {\cite{LS}}
the construction of such a black hole is given following
the observation in {\cite{AO}} that the 3D rotating BTZ black
hole\cite{btz} might be
interpreted as an electrically charged 2D black hole.

Motivated partly by the results of \cite{BY} in showing the
connection between the stability of the 4D black holes with
the entropy at finite distance, we study in this paper various
types of 2D black holes. First, we explicitly show that
the thermodynamic
relations are well-defined for the observations made
from a finite distance.
We then investigate the thermodynamics of a charged 2D black
hole with asymptotic properties similar to the ones
mentioned above.
In comparing our results with those from other methods,
we find that one has to make appropriate gauge transformations of
the gauge potentials in order to obtain a consistent value of
the entropy.

We also study the
specific heat of these solutions.
An analysis taking into account the definition
of the observed temperature shows that specific heat is
positive for all the observers.

Let us now start with the discussions of the class
of 2D black holes
obtained in \cite{CM}. These emerge as the solutions
of the action:
\be \label{jt}
{\cal I} = -\int_{\cal M} \sqrt{g}
\exp{(-2 \phi)}  \left[ R +
\frac{8 k}{k - 1}(\nabla \phi )^2 + \lambda^2 \right] -
2\int_{\partial {\cal M}} \exp(-2\phi) K,
\ee
where $K$ is the trace of the second fundamental form, $\partial
{\cal M} $ is the boundary of $\cal M$ and
$k$ is a parameter taking values $\mid k \mid \le 1$.
It reduces to the Jackiw-Teitelboim action \cite{jackiw}
for $k=0$. Equations of motion ensuing from (\ref{jt}) are:
\bea \label{cmsoln1}
R + \frac{8k}{k - 1} (\nabla^2\phi - (\nabla\phi)^2) + \lambda^2
& = & 0 \\ \label{cmsoln2}
R_{ab} -\frac{1}{2}g_{ab} R - \frac{1}{2} g_{ab}\lambda^2
+ 2 \nabla_a \nabla_b \phi & - & 4 \frac{1 + k}{1 - k} \nabla_a \phi
\nabla_b \phi \nonumber \\
& - &2 g_{ab} \left[ \nabla^2 \phi - \frac{2}{1 -k}
(\nabla\phi)^2 \right]  =  0.
\eea
These possess the following exact solutions:
\be \label{exsol1}
ds^2 = -\sinh^2(\kappa \sigma)\cosh^{2k} (\kappa\sigma)dt^2 +
d\sigma^2
\ee
\be \label{exsol2}
\exp{(-2\phi)} = \exp{(-2\phi_0)} \cosh^{1 - k}(\kappa \sigma),
\ee
where $\kappa = \frac{\lambda}{\sqrt{2(1 - k)}}$.
These solutions are everywhere regular for any value of $k$
and have a horizon at $ \sigma = 0 $. They
asymptote to the anti-de Sitter background with
a linear dilaton for $ \sigma \rightarrow \infty $.
Also note that the solution
(\ref{exsol1})--(\ref{exsol2}) describes, for $ k =
-1 $, the usual asymptotically flat stringy dilatonic black
hole\cite{witten}.
Thermodynamics in this special case is dealt with at length in
{\cite{GP}}. We shall
study the thermodynamics of the solutions
(\ref{exsol1})--(\ref{exsol2}) for
generic $k$.

The discussion of thermodynamics  begins with the definition
of free energy:
\be
\cal F = \frac{\cal I}{\beta},
\ee
where $\cal I $ is the Euclideanized
action evaluated for the metric of the
space-time under consideration and $\beta $ is the inverse
temperature. Therefore we will start with the evaluation of the
action (\ref{jt}). Since we are
concerned with the thermodynamics of the black holes at a finite
spatial separation, we have to evaluate the
action with the boundary
contribution  on a spacelike slice. This is
feasible
since the solutions (\ref{exsol1}) admit a Killing
vector $ k^a = \left( \frac{\partial}{\partial t} \right)^a $.
The proper periodicity of the
Euclideanized time at a fixed value of the spatial coordinate
is interpreted as the local temperature $ T_{w} =
\beta^{-1} $. The conserved dilaton current $ j_a \equiv
\epsilon_{a}^{b} \nabla_b \exp(-2\phi) $ defines another
thermodynamic potential
\be \label{dildef}
{\cal D} = \int_{\Sigma} j
\ee
where $\Sigma $ is a spacelike hypersuface bounded by the wall
of the box. Here we note a direct consequence of the non-minimal
coupling of the dilaton in two dimensions.
In four dimensions, the dilaton
field can be decoupled from the Einstein term by
rescaling the metric.
As a result
it becomes the part of a general matter action and does not
affect the black hole thermodynamics \cite{GP,KOP}.
Also, the
dilaton charge $\cal D$ in equation(\ref{dildef})
 is essentially the value of the dilaton
field $\exp(-2\phi)$ on the boundary, which is a scalar.
Consequently this quantity is a measurable one,
in contrast to the
coordinates which parametrize it\cite{GP}. We will
therefore treat $\cal
F$ as a function of the two
thermodynamic quantities $ T_{w}$ and $\cal D $. Then
by the first law of thermodynamics,
one can define the entropy $ S $ and the dilaton potential
$\psi $ as:
\be
S  =  - \left[\frac{\partial {\cal F}}{\partial T_{w}} \right]_{\cal
D} , \,\,\,
\,\,\,\,\,\,\,\,\,\,\,\,\,\,\,\,\,\,
\psi  =  - \left[ \frac{\partial {\cal F}}{\partial {\cal D}}
\right]_{T_{w}} .
\ee
But it is not the
Helmholtz free energy, rather its Legendre transform, $ {\cal E} =
{\cal F} + ST_{w} $,
that defines the {\em non-available} energy. This
corresponds to the mass of the space-time,
provided it exsists, as the limiting value of the
difference between the energies for the black hole and
its asymptotic background solution
at spatial infinity.

Using the dilaton equation(\ref{cmsoln1}),
the action (\ref{jt}) can be evaluated for a generic $k$
to be
\be \label{bdry}
{\cal I} = -2 \int_{\partial {\cal M}} \exp({-2 \phi}) \left( K
- \frac{4k}{k-1} n^{a} \nabla_{a} \phi \right),
\ee
where $ n^a $ is the unit
outward normal on $\partial{\cal M} $. For our choice of
boundary, $n^{\mu} = (0, \frac{1}{\sqrt{g_{11}}})$, $ K =
\frac{\partial_1 \ln \sqrt{\mid g_{00} \mid}}{\sqrt{g_{11}}}$
and the action has the form
\be \label{act}
{\cal I} = -\int_{\partial{\cal M}} \sqrt{\frac{1}{g_{11}}}
\exp(-2\phi) \left(
\frac{1}{2} \frac{\partial_1
g_{00}}{g_{00}} - \frac{4k}{k-1} \partial_1 \phi\right)
\ee
Then by defining $ x = \kappa \sigma $,
the Helmholtz free energy for the solution (\ref{exsol1}) becomes
\bea \label{cmfe1}
{\cal F}  & = & {\cal I} T_{w} \\ \label{cmfe2}
& = & - 2\kappa \frac{T_w}{T_c} \exp(-2\phi_{0}) \left[
\cosh^2 x - k \sinh^2 x \right]
\eea
where $T_c$ is the proper periodicity of the Euclidean time
at the horizon and is related to $T_w$ by
\be
T_w = \frac{T_c}{\sqrt{g_{00}}}
\ee
The Euclideanized metric corresponding to
(\ref{exsol1}) has a conical
singularity as $\sigma \rightarrow 0$, unless
$\tau \equiv i t$ has a periodicity
$ T_c = \frac{\kappa}{2 \pi} $.
Thus,
\be \label{twtc}
T_w = \frac{T_c}{\sinh x \cosh^k x}
\ee
and the dilaton charge for this black hole is
\be \label{dil}
{\cal D} =  \exp(-2\phi_{0} ) \cosh^{1 - k}x.
\ee
Then using (\ref{cmfe2}) and
(\ref{dil}) we find
\be \label{cment}
{\cal F}  =  -2 \kappa{\cal D} \left( \coth x - k\tanh x
\right).
\ee
In (\ref{cment}) we have eliminated
the constant $\exp(-2\phi_0)$ in favor
of the dilaton charge $\cal D$, the basic principle being, that
one should not keep arbitrary parameters in the description of the
thermodynamic quantities except those which appear
in the action itself. But the coordinate $x$ is kept as an
implicit variable defined by (\ref{twtc})\cite{GP}.
Since both $\cal F $ and $ T_{w} $ depend implicitly on $
x$, we can write $ S $ as
\be  \label{entdef}
S  =  - \left[\frac{\partial {\cal F}}{\partial x} \right]_{\cal
D} \left[ \frac{dT_{w}}{dx} \right]^{-1}.
\ee
This yields
\be \label{cmen}
 S  =  4 \pi {\cal D} \cosh^{k-1}x
 =  4 \pi \exp(-2 \phi_{0}).
\ee
The black hole energy as defined by the Legendre transform
of $\cal F$ is given by
\be \label{ngen}
{\cal E}_{BH} = -2(1- k) \kappa {\cal D} \tanh x
\ee
We observe that all the thermodynamic
quantities listed above go over to those for the string
black hole by choosing $ k = -1 $\cite{GP}. In fact, the
entropy is a constant
of  $x$ for all values of $k$ including
the case of string black hole.
We however notice some important differences
between $k= -1$ and $k\neq -1$ situations.
Unlike the $k= -1 $ case
the solutions (\ref{exsol1})--(\ref{exsol2}) asymptote
 to the anti-de Sitter
(AdS) linear dilaton vacuum for general $k$. As a result $T_w$
vanishes asymptotically as $\exp[-(k+1)x]$ while $\cal D$ goes
to infinity as $\exp[(k+1)x]$.
The energy of the black hole is
to be computed with reference to the AdS linear dilaton vacuum
defined by
\bea \label{ads}
ds^2 & = & \exp(2(k + 1) \kappa \sigma) dt^2 + d \sigma^2 ,\\
\phi & = & \phi_0 + \frac{1}{2} (k - 1) \kappa \sigma.
\eea
The free energy (\ref{cmfe1}) becomes
\be
{\cal F}_{AdS} = -2\lambda (1 - k) {\cal D},
\ee
which implies $ S_{AdS} = 0 $ and $ {\cal E}_{AdS} = -2\lambda (1
- k) {\cal D} $. Then defining $ M \equiv {\cal E}_{BH} - {\cal
E}_{AdS} $ we obtain
\be \label{cmmass}
M = 2 \kappa {\cal D} (1 - k) [1 - \tanh x].
\ee

An analogue of (\ref{first}) at finite $x$ was written
in \cite{GP} for $k=-1$ and has the form
\be
S = \frac{M}{T_c} \left(1 - \frac{M}{16 \pi {\cal D} T_c }\right).
\ee
It can be verified that the above equation generalizes for general
$k$ to,
\be \label{SM}
S  = 4\pi {\cal D} \left[ \frac{M}{2 \pi {\cal D} T_c (1
-k ) }\left(1 - \frac{M}{8 \pi {\cal D} T_c (1-k )}\right)
\right]^{(1-k)/2}.
\ee
Note that unlike the case of asymptotically flat metric, the
quantity $M$ for a general $k$ vanishes in the limit
$ x \rightarrow \infty$ by the Tolman redshift factor as
$M \sim M_{ADM} \exp{[-(k+1) x]} $, where $ M_{ADM} = (1 -
k)\frac{\kappa}{2} \exp (-2 \phi_0)$ is the
ADM-mass of the black hole. Equation (\ref{SM}) is one of the main
results of this paper.
Also, one can verify that, in the asymptotic limit,
\be
S T_c = \frac{2 M_{ADM}}{1 - k},
\ee
which is precisely the relation given in \cite{CM}.

We now investigate the thermodynamics for the charged version of
the $k = 0$ black hole. For $k=-1$ the charged
black hole solution
and its thermodynamics is discussed in \cite{nappi} and \cite{GP}
respectively. For $k=0$, gauge fields were
introduced in \cite{LS} through the dimensional compactification
of a three dimensional string effective action using a
suggestion in \cite{AO}. The 2D action in this case has the form,
\be \label{action32d}
{\cal I} = -\int_{\cal M_{\rm 2}} \sqrt{g} \exp{(-2 \Phi)} [ R +
 2 \lambda^2 -\frac{1}{4}\exp(-4\Phi) F^2] - 2\int_{\partial
{\cal M}} \exp(-2\Phi) K,
\ee
where now $ \Phi $ is a scalar field coming from the
compactification and plays the role of  dilaton for the 2D
action. Action
(\ref{action32d}) describes the Jackiw-Teitelboim theory with a gauge
field. The equations of motion ensuing from this action
are
\bea \label{lseom1}
R_{ab} + 2\nabla_a \nabla_b \Phi - 4\nabla_a \Phi \nabla_b \Phi
 +  \frac{1}{2} \exp(-4\Phi)F_a^c F_{cb} & - & \nonumber \\
g_{ab} \left[ \frac{1}{2} R + \lambda^2 + 2\nabla^2\Phi - 4(\nabla
\Phi)^2 - \frac{1}{8} \exp(-4\Phi) F^2 \right] & = & 0, \\
\label{lseomi}
R + 2\lambda^2 -\frac{3}{4} \exp(-4 \Phi) F^2 & = & 0, \\
\label{lseom2}
\partial_a \left(\sqrt{g}\exp (-6 \Phi) F^{ab} \right) & = & 0.
\eea
These possess the solution
\bea \label{lsbh1}
ds^2 & = & -( M - \lambda^2 r^2 - \frac{J^2}{4 r^2}) dt^2 +
( M - \lambda^2 r^2 - \frac{J^2}{4 r^2} )^{-1} dr^2, \\ \label{lsbh}
A_0 &= &- \frac{J}{2 r^2}, \\ \label{lsbh2}
\exp(- 2 \Phi) &= &r.
\eea
The parameter $J$ in this solution gives charge to this
black hole.
The metric has a curvature singularity at $ r = 0 $
for nonvanishing $J$ as is seen from the Ricci scalar
\be
R = - 2 \lambda^2 - \frac{ 3 J^2 }{2 r^4}.
\ee
It also goes asymptotically, $r \rightarrow \infty$, to the
anti-de Sitter Space-time.
Now we study the thermodynamics of these black
hole solutions for observations done from finite distances.
In this case, the use of the equations of motion
(\ref{lseom1})--(\ref{lseom2})
implies the following value of the classical action:
\be \label{lsact}
{\cal I} = -\int_{\partial{\cal M}} \left[ n^a F_{a b} A^b
\exp(-6\Phi) + 2 K \exp(-2 \Phi) \right].
\ee
To discuss the thermodynamics, we rewrite the solution
(\ref{lsbh1})--(\ref{lsbh2})
in the non-extremal case, $ M^2 > \lambda^2 J^2 $,
in the coordinates:
\be \label{transf}
r^2 = \frac{M + \sqrt{M^2 - \lambda^2 J^2}\cosh
2\lambda\rho}{2\lambda^2}
\ee
by exploiting the fact that  it admits a timelike Killing vector.
Then we find,
\be
ds^2 = -{\cal G}(\rho) dt^2 + d\rho^2 \\
\ee
where
\be
{\cal G}(\rho) = \frac{1}{2} \frac{(M^2 - \lambda^2 J^2)
\sinh^2 2\lambda\rho}{M + \sqrt{M^2 - \lambda^2 J^2}\cosh
2\lambda \rho} .
\ee
$A_0$ and $\exp(-2\Phi)$ are still given by
(\ref{lsbh})--(\ref{lsbh2}) with $r$ replaced from (\ref{transf}).
In these coordinates the horizon is at $\rho=0$. The free energy is
obtained by the evaluation of (\ref{lsact}). We note
however that there is
an ambiguity in the evaluation of (\ref{lsact})
due to the freedom of a
constant shift in the gauge potential: $A_a\rightarrow A_a +
const.$, in the equations of motion. Constant shifts have been
applied earlier\cite{KOP,GibHaw1}
 in the evaluation of the classical actions in
order to avoid divergence in the gauge potential at the horizon.
In our case, on the other hand, $A_a$ is well-defined at $\rho=0$.
But as we will see later, this shift is needed in order to
show the consistency of the present method of computations
with the Noether's charge prescription \cite{wald}.

Once again the temperature is given by the periodicity of the
proper time in a local inertial frame around $\rho$ and satisfies
the relation:
\be \label{temp}
T_w = \sqrt{2} T_c \frac{[M +
\sqrt{M^2 - \lambda^2 J^2}\cosh 2x]^{1/2}}
{\sqrt{M^2 - \lambda^2 J^2}\sinh 2x}.
\ee
where $x = \lambda \rho$ and
\be
T_c = \frac{\sqrt{2} \lambda}{2 \pi} \frac{\sqrt{M^2 - \lambda^2
J^2}}{(M + \sqrt{M^2 - \lambda^2 J^2})^{1/2}}
\ee
is the proper periodicity at the horizon.
The dilaton charge is now given by
\be \label{lsdil}
{\cal D} = \left[ \frac{M}{2\lambda^2} (1 + \xi \cosh 2x)
\right]^{1/2},
\ee
where
$\xi = \sqrt{1 - \left(\frac{\lambda J}{M}\right)^2}$  with $\xi^2
 > 0$.
Then one can evaluate (\ref{lsact}), with a shift in the gauge
potential $ A_\mu \rightarrow A_\mu(\rho) - A_\mu (\rho = 0) $,
and the free energy is
\be \label{lsfe}
{\cal F} = -2 \lambda {\cal D} \coth x.
\ee
The form of equation (\ref{lsfe}) needs some
qualifications.  As in (\ref{cment}), an
implicit variable $x$ has been used in writing them.
However, the thermodynamic variables are only the dilaton
charge($\cal D$), temperature$(T_w)$ and the electric
charge$(Q)$, defined as
$ Q = -\frac{1}{2} \exp(-6\Phi) \epsilon_{ab} F^{ab}$ at the
boundary.
To show that
the free energy can be written purely in terms of $\lambda $ and
thermodynamic variables
$T_w$, $Q$ and ${\cal D}$, it suffices to record the following
relations:
\be
\frac{Q}{{\cal D}^3} = 2\lambda^2\xi \frac{\sqrt{1 - \xi^2}\sinh
2x}{( 1 + \xi \cosh 2x)^2},
\ee
and
\be
\xi = \frac{\pi^2 T_{w}^2 \sinh^2{2 x} - \lambda^2}{\lambda^2
\cosh 2x - \pi^2 T_{w}^2 \sinh^2 2x}.
\ee
Since $\cal F$ in (\ref{lsfe}) does not depend explicitly
on $Q$ and $\xi$,
 entropy is once again computed using equation
(\ref{entdef}) and can be
written as
\be \label{lsent}
S  = - \frac{4 \lambda {\cal D} \coth x}{T_w}
\left[ \frac{\xi (1 + \xi \cosh 2x )}{1 - \xi^2 -
(1 + \xi \cosh 2x)^2}
\right].
\ee
The consistency of this procedure is provided by the
fact that
\be
\frac{d\xi}{dT_w} \equiv \frac{\partial \xi}{\partial T_w} +
\frac{\partial \xi}{\partial x}\left( \frac{dT_w}{dx}
\right)^{-1} = 0.
\ee
As a result, in differentiating with respect to $T_w$ and $x$,
$\xi$ is taken as a constant.
In the same manner as above, $S$ can also be thought to
be a function of thermodynamic variables only.
In the $ x \rightarrow \infty $ limit we find
\be \label{lsenasy}
\lim_{x \rightarrow \infty} S =
 \frac{2\sqrt{2}\pi}{\lambda}\left[ M +
\sqrt{M^2 - \lambda^2 J^2} \right]^{1/2}.
\ee
Recently, the black hole entropy
for asymptotic observers
has been evaluated by different
methods, includuing the Noether's charge
prescription\cite{wald,rcm}.
In the case of 2D black holes, this simply leads to $ S =
4\pi\exp(-2 \Phi) $ evaluated on the horizon. This result matches
with the ones derived in (\ref{lsenasy}). We would like to
remind the reader of the crucial role of the gauge choice in
deriving (\ref{lsenasy}) for this comparison.

We now come to the stability analysis of the
black holes through the evaluation of the specific heat.
The space-time is
thermodynamically stable to radiation provided the specific heat
is positive.
It is noted that, at least in those cases, where $S$ and $T_w$
are asymptotically constants of $x$, say, $S_0$ and $T_0$,
respectively, an equation of the type (\ref{first})
is satisfied and the specific heat,
\be \label{spec}
{\cal C}  =  T_0 \left( \frac{dS_0}{dT_0} \right), \\
\ee
is negative, viz,  $-S_0$.
For the case under consideration, however, it is
 naive to conclude from this that the black
hole is unstable.

We now compute the specific heat in the present formulation and
show the stability of the black hole solutions. The specific
heat is now given by the formula:
\be \label{spht}
{\cal C_D} \equiv T_w \left[ \left( \frac{\partial
S}{\partial T_w }\right)_{x, \cal D} + \left(
\frac{\partial S}{\partial
x} \right)_{T_w,\cal D}
\left(\frac{dT_w}{dx}\right)_{\cal D}^{-1} \right].
\ee
For the uncharged black holes (\ref{exsol1})--(\ref{exsol2}),
using the entropy (\ref{cmen}), we obtain the specific heat as:
\be
{\cal C}_{\cal D} = 4 \pi (1 - k)  \frac{\exp(-2\phi_0)}{k +
\coth^2 x},
\ee
which is positive for all $ \mid k \mid < 1$.
Therefore one concludes
that these black holes are stable.
For $k=-1$, on the other hand, $\cal C_D$ is infinite
asymptotically. This coforms to the observations made
earlier\cite{trivedi}.

For the charged black hole, the specific heat
for a constant ${\cal D}$ and $Q$ is found to be:
\bea
{\cal C}_{{\cal D}Q} & = & \frac{8 \pi
\xi}{\lambda}\sqrt{\frac{M}{2}(1+\xi)} \frac{\cosh^2 x (1 + \xi
\cosh 2x)}{[(1 - \xi^2) - (1 + \xi \cosh 2x)^2]^2}
\left[\frac{{}^{}}{{}^{}} ( 1 - \xi^2) \right. \nonumber \\
& - &(1 + \xi \cosh 2x)^2
 +  2\xi (1 + \xi \cosh 2x) \nonumber \\
& - & \left. 2\xi^2 \sinh^2 2x \frac{(1 -
\xi^2) + (1 + \xi \cosh 2x)^2}{(1 - \xi^2) - (1 + \xi \cosh
2x)^2} \right],
\eea
where once again we have used the constancy of $\xi$.

We have plotted ${\cal C_{D{\rm Q}}}$ as a function of $x$
in Fig.1 for certain values of
$\xi$ and found that it is positive throughout.
Its asymptotic value is
 same as that of entropy, $S$ in (\ref{lsenasy}). In the other limit,
$ x \rightarrow 0$, the specific heat vanishes as $\sim
x^2$. It is now interesting to note
that for $x$ close to zero we also have ${\cal
C}_{\cal D \rm Q} \sim T{_w}^2$. As is known
that a power law dependence of specific heat on
temperature is a signature of the presence of massless modes
in a theory. Its significance in our context, in the light of
masslessness of the dilaton, should be interesting to analyze.

To summarize, in this paper we have investigated black hole
thermodynamics at finite distance for several types of black
holes. We have also argued for their thermodynamic stability by
calculating the specific heat. We also point out the role of the
gauge freedom in the consistent evaluation of thermodynamic
potentials. It will be interesting to further generalize our
results in many directions. First, it has been already pointed
out that the action (\ref{jt}) has a duality
symmetry for general $k$
\cite{CM1}. It will be of interest to find out the nature of the
thermodynamics for the dual black holes and compare it with the
present results. In this regard, the duality invariance of
thermodynamic quantitites have been shown for string case
ealier\cite{horow}.
Whether these results are still valid for
observations at finite distances is worth addressing.
Secondly, the surface terms have been written down earlier
for the higher
curvature gravity theories. In the context of string theories,
since the black hole solution is already known to all-orders,
one can also study their thermodynamics to higher orders in this
method.
Whether there is a way to address the all-order (in $\alpha'$)
thermodynamics for string theories is an open question.
Moreover, one can possibly
generalize our results to include the dilaton potential. We
expect that in that case the nature of specific heat will differ
from the results presented here, the reason being the absence of
a massless mode.
Finally, the results for the charged black hole
derived in this paper apply only to
non-extremal black holes. Although the expression for
entropy (\ref{lsent})
has a well-defined limit as $\xi\rightarrow 0$,
this is not quite the correct value of the
asymptotic entropy for the extremal black hole.
The transformation(\ref{transf}) is valid
only if $\xi > 0 $. In fact,
as has recently been advocated by Hawking et al in \cite{HHR},
the extremal black hole is
thermodynamically a different object than the non-extremal one.
The extremal black hole has to be treated separately.
It will be interesting to see how the considerations
of \cite{HHR} translate
to the cases treated here.

\section*{Acknowledgement}
It is a pleasure to thank S P Khastgir for many useful
discussions during this work.


\newpage
\section*{Figure Caption}
{\bf Fig 1.} Plot of specific heat ($\cal C_{D\rm Q}$) vs.
coordinate($x$) for the black hole solution (\ref{lsbh1}) for
$M = \lambda = 1$. Curves are labelled by different values of
$\xi$. The origin on the x-axis is shifted. The curves start at
$x=0$.

\end{document}